\documentclass[amsmath,amssymb,preprintnumbers,nofootinbib,prd,a4paper,twocolumn]{revtex4-1}
\pdfoutput=1
\usepackage{amsthm}
\usepackage{amsmath}
\usepackage{graphicx}
\usepackage{bbold}
\usepackage{color}
\usepackage{subfigure}
\usepackage{multirow}
\usepackage{graphicx,bm}

\usepackage{dcolumn}
\usepackage{bm}
\usepackage{slashed}
\usepackage{epsfig}
\usepackage{hyperref}
\usepackage{verbatim}

\usepackage[utf8]{inputenc}
\usepackage[T1]{fontenc}
\usepackage[normalem]{ulem}

\newcommand{\beq}{\begin{eqnarray}}
\newcommand{\eeq}{\end{eqnarray}}

\newcommand{\bmp}{\noindent\begin{minipage}{16cm}}
\newcommand{\emp}{\end{minipage}\vskip 7mm} 

\newcommand{\GeV}{\mbox{ ${\mathrm{GeV}}$}}

\newcommand{\be}{\begin{eqnarray}}
\newcommand{\ee}{\end{eqnarray}}

\begin{document}
\title{Higgs boson emerging from the dark}

\author{Chengfeng {\sc Cai}}
\affiliation{School of Physics, Sun Yat-Sen University, Guangzhou 510275, China}
\author{Giacomo {\sc Cacciapaglia}}
\email{g.cacciapaglia@ipnl.in2p3.fr}
\affiliation{Institut de Physique des 2 Infinis (IP2I),
CNRS/IN2P3, UMR5822, 69622 Villeurbanne, France}
\affiliation{Universit\' e de Lyon, Universit\' e Claude Bernard Lyon 1, 69001 Lyon, France}
\author{Martin {\sc Rosenlyst}}
\email{rosenlyst@cp3.sdu.dk}
\affiliation{CP$^3$-Origins, University of Southern Denmark, Campusvej 55, DK-5230 Odense M, Denmark}
\author{Hong-Hao {\sc Zhang}}
\email{zhh98@mail.sysu.edu.cn}
\affiliation{School of Physics, Sun Yat-Sen University, Guangzhou 510275, China}
\author{Mads T. {\sc Frandsen}}
\email{frandsen@cp3.sdu.dk}
\affiliation{CP$^3$-Origins, University of Southern Denmark, Campusvej 55, DK-5230 Odense M, Denmark}


\begin{abstract}
We propose a new non-thermal mechanism of dark matter production based on vacuum misalignment. A global $X$-charge asymmetry is generated at high temperatures, under which both the will-be Higgs and the dark matter are charged. At lower energies, the vacuum changes alignment and breaks the $U(1)_X$, leading to the emergence of the Higgs and of a fraction of charge asymmetry stored in the stable dark matter relic. This mechanism can be present in a wide variety of models based on vacuum misalignment, and we demonstrate it in a composite Higgs template model, where all the necessary ingredients are naturally present. A light pseudo-scalar $\eta$ is always predicted, with interesting implications for cosmology, future supernova observations and exotic $Z \to \gamma \eta$ decays.
\end{abstract}

\maketitle


The presence of Dark Matter (DM) in the Universe is arguably one of the most important mysteries in our knowledge of the physical world. We have compelling evidence in cosmological and astrophysical observations that the majority of the matter density in the whole Universe~\cite{Aghanim:2018eyx} and around galaxies and galaxy clusters is of non-baryonic nature. Nevertheless, no particle candidate exists within the Standard Model (SM) of particle physics to fill this gap.
The most popular paradigm has been the WIMP one, postulating the presence of a new Weakly Interacting Massive Particle beyond the SM. As this paradigm is currently challenged by the non-observation of a signal in direct detection experiments~\cite{Aprile:2018dbl}, many new mechanisms have been recently proposed: Asymmetric DM~\cite{Nussinov:1985xr,Kaplan:2009ag}, freezing-in FIMPs (`F' for feebly)~\cite{Hall:2009bx}, $2\to3$ annihilating SIMPs (`S' for strongly)~\cite{Carlson:1992fn,Hochberg:2014dra,Hansen:2015yaa}, to name a few.

In this letter we propose a new mechanism for non-thermal DM production based on vacuum misalignment, in models where the Higgs arises as a pseudo-Goldstone boson from a spontaneously broken global symmetry. This class of models includes composite Higgs models~\cite{Kaplan:1983fs}, holographic extra dimensions~\cite{Contino:2003ve,Hosotani:2005nz}, Little Higgs~\cite{ArkaniHamed:2001nc,ArkaniHamed:2002qx}, Twin Higgs~\cite{Chacko:2005pe} and elementary Goldstone Higgs models~\cite{Alanne:2014kea}.
Our main starting point is the fact that the vacuum of the theory, in general, depends on the temperature of the Universe. Thus, its structure today at zero temperature (where the misaligned Higgs vacuum is needed) and at the global symmetry breaking scale is likely to be different.
Within this set-up, we propose that a DM relic density may be asymmetrically produced via a charge that is preserved only in the high-temperature vacuum. At low temperatures, the charge is broken  and the fraction of asymmetry stored in $\mathbb{Z}_2$--odd states remains as DM density. The main advantage is that, while the non-thermal DM production is due to an asymmetry, the low energy DM candidate can decouple from the SM thus avoiding conflict with direct detection data.
The mechanism we propose requires the following key ingredients:
\begin{itemize}
\item[i)] an exact $\mathbb{Z}_2$ symmetry that keeps a DM candidate stable;
\item[ii)] right below the global symmetry breaking phase transition at $T_{\rm HL}$, the vacuum of the theory is an essentially Higgsless (HL) phase\footnote{We refer to this as a Higgsless phase even though a state with the quantum numbers of the isosinglet Higgs is present because this state is expected very heavy at the transition.}, where the electroweak symmetry is broken at a scale $f \gg v_{\rm SM} = 246$~GeV, and a global $U(1)_X$ symmetry remains unbroken;
\item[iii)] at $T_{\rm HL}$, an asymmetry is generated in - or transferred to - the $U(1)_X$ charged states, some of which are also odd under $\mathbb{Z}_2$;
\item[iv)] at $T_\ast < T_{\rm HL}$, the vacuum starts rotating away from the TC vacuum, and $U(1)_X$ is spontaneously broken;
\item[v)] at $T \approx 0$, the theory settles on the standard pseudo-Goldstone Higgs vacuum, where the misalignment reproduces the electroweak (EW) scale $v_{\rm SM}$.
\end{itemize}
In this process, the fraction of asymmetry stored in $\mathbb{Z}_2$--odd states at $T = T_\ast$ will survive as DM density as long as such states are decoupled from the SM thermal bath. Furthermore, as we will see in an explicit example, the pseudo-Goldstone Higgs emerges from the $\mathbb{Z}_2$--even states charged under the $U(1)_X$ while the vacuum rotates away from the HL vacuum.

To demonstrate how the mechanism works, we will consider
models of composite Higgs with vacuum misalignment, which can fulfil all the above requirements. For concreteness, we will focus on models based on an underlying gauge-fermion description, for which the symmetry breaking patterns are known~\cite{Witten:1983tx,Kosower:1984aw}: the minimal cosets with a Higgs candidate and custodial symmetry are $SU(4)/Sp(4)$~\cite{Galloway:2010bp}, $SU(5)/SO(5)$~\cite{Dugan:1984hq} and $SU(4)\times SU(4)/SU(4)$~\cite{Ma:2015gra}. A $\mathbb{Z}_2$ symmetry is already present in the latter case~\cite{Ma:2017vzm}, while the other two cases can be easily extended to a $SU(6)$ symmetry~\cite{Cai:2018tet,Cacciapaglia:2019ixa}.
A global $U(1)_X$ in the HL vacuum has already been used to define a DM candidate in a $SU(4)/Sp(4)$ Technicolor-like theory in Ref.~\cite{Ryttov:2008xe} (the connection to the composite Goldstone Higgs vacuum has been studied in Ref.~\cite{Cacciapaglia:2014uja}). We have checked that a $U(1)_X$ can also be defined in the $SU(4)\times SU(4)/SU(4)$ coset (but not in $SU(5)/SO(5)$). Note that the above features can also be found in other cosets that do not have a simple gauge-fermion underlying description, like the models in Refs~\cite{Ballesteros:2017xeg,Balkin:2017aep}, and can also be found in elementary realisations.~\footnote{In all models with a $\mathbb{Z}_2$ symmetry, the pNGB dark matter relic density could also be obtained by thermal freeze-out~\cite{Frigerio:2012uc}.} Our proposal is therefore rather general.

To better understand the workings of this mechanism, we need to recall some basic information about modern composite Higgs models~\cite{Contino:2010rs,Bellazzini:2014yua,Panico:2015jxa} based on vacuum misalignment: a Higgs-like boson arises as a composite pseudo-Nambu-Goldstone boson (pNGB) from the breaking of a global symmetry $\mathcal{G}$ to $\mathcal{H}$. The model is such that an alignment exists where $\mathcal{H}$ contains the EW gauge symmetry $SU(2)_L \times U(1)_Y$.
This alignment, however, is not stable as an explicit breaking of $\mathcal{G}$ exists in the form of gauge interactions, top couplings to the strong sector, and current masses for the confining fermions. These terms are responsible for generating a vacuum expectation value for the Higgs, which corresponds to a misalignment of the vacuum. We will describe this by an angle, $\sin \theta = v/f$~\cite{Kaplan:1983fs}. At $T\approx 0$, we need $v(0) \equiv v_{\rm SM} =246$~GeV to reproduce the SM at low energy. The decay constant of the pNGBs (including the Higgs), $f$, is fixed by the degree of fine-tuning in the zero-temperature potential:
typically, $\sin \theta \lesssim 0.2$ from electroweak precision measurements~\cite{Agashe:2006at,Barbieri:2012tu,Grojean:2013qca}, thus fixing $f \gtrsim 1.3$~TeV, even though smaller scales may also be allowed~\cite{Ghosh:2015wiz,BuarqueFranzosi:2018eaj}.  We stress that $f$ is a fixed scale, only depending on the confinement of the underlying strong dynamics, while it's the value of $v(T)$ at the minimum of the potential that varies with temperature.
As we assume that the vacuum is only misaligned along the Higgs direction, the coset structure can be schematically represented by a $n_f \times n_f$ matrix:
\beq
\left( \begin{array}{c|c}
\mathcal{G}_0/\mathcal{H}_0 & \begin{array}{c} \mathbb{Z}_2\mbox{--odd} \\ \mbox{pNGBs} \end{array} \\ \hline
\begin{array}{c} \mathbb{Z}_2\mbox{--odd} \\ \mbox{pNGBs} \end{array} & \begin{array}{c} \mathbb{Z}_2\mbox{--even} \\ \mbox{pNGBs} \end{array}
\end{array} \right)\,,
\eeq
where $\mathcal{G}_0/\mathcal{H}_0$ is one of the two minimal cosets listed before. The origin of the parity can be easily understood in terms of the underlying fermions $\psi^i$, $i = 1, \dots n_f$, that condense: the $\mathbb{Z}_2$ parity flips sign to the $\psi^{5, \dots n_f}$ fermions that do not participate to the minimal coset, while the $U(1)_X$ will materialise as a phase acting on the fermions $\psi^{1, \dots 4}$ in the minimal cosets. This assignment also explains why we expect $\mathbb{Z}_2$--odd pNGBs carrying $U(1)_X$ charges in the off-diagonal block.  An explicit vector like mass term breaks the would-be $U(1)_X$ charge of the fermions $\psi_{5,6}$ in the HL vacuum.
For concreteness, we will use the $SU(6)/Sp(6)$ model~\cite{Cai:2018tet} as a template.~\footnote{Note that in Ref.~\cite{Cai:2018tet} the authors focus on a scenario where a $U(1)_{\rm DM}$ is preserved in the Higgs vacuum, case that is disfavoured by direct detection.}  We assume 4 Weyl fermions are arranged in $SU(2)_W$ doublets, $\psi_{1,2}$ and $\psi_{5,6}$ and two in $SU(2)_W$ singlets $\psi_{3,4}$. We list in Table~\ref{tab:su6sp6} the quantum numbers of the pNGBs in the $\theta=0$ vacuum and in the HL vacuum ($\theta = \pi/2$). Note how the will-be Higgs $h$ and the singlet $\eta$ form the only $\mathbb{Z}_2$--even state charged under $U(1)_X$ in the HL vacuum. In the Higgs vacuum, the DM candidate is a real scalar, and thus the absence of a coupling to the $Z$ ensures that direct detection bounds are avoidable.

\begin{table}[tb]
\begin{center}
\begin{tabular}{c|c|c|c}
\hline\hline
    & $\begin{array}{c} \mbox{Higgs vacuum} \\ \theta \sim 0 \end{array}$ & $\begin{array}{c} \mbox{HL vacuum} \\ \theta = \pi/2 \end{array}$ & $q_X$\\
\hline\hline
$\displaystyle \frac{\mathcal{G}_0}{\mathcal{H}_0}$ &  $\begin{array}{c} H_1 = 2_{1/2} \\ \eta = 1_0 \end{array}$ & $\begin{array}{c} \phi_X = (h+i \eta)/\sqrt{2} \\ \omega^\pm\,, \;\; z^0 \end{array}$ & $\begin{array}{c} 1 \\ 0 \end{array}$\\
\hline
$\begin{array}{c} \mathbb{Z}_2\mbox{--odd} \\ \mbox{pNGBs} \end{array}$ & $\begin{array}{c} H_2 = 2_{1/2} \\ \Delta = 3_0 \\  \varphi = 1_0 \end{array}$ & $\begin{array}{c} \Theta_{1} = - H_2^0 + \frac{\Delta_0 + i \varphi_0}{\sqrt{2}} \\ \Theta_{2} =  (H_2^0)^\ast + \frac{\Delta_0 - i \varphi_0}{\sqrt{2}} \\ \Theta^-_{1} = \Delta^- - H^- \\ \Theta^+_{2} = \Delta^+ + H^+ \\ + \mbox{c.c.} \end{array}$ & $\displaystyle \frac{1}{2}$\\
\hline
$\begin{array}{c} \mathbb{Z}_2\mbox{--even} \\ \mbox{pNGBs} \end{array}$ & $\eta' = 1_0$ & $\eta'$ & $0$\\
\hline\hline
\end{tabular} \end{center}
\caption{pNGBs in the template $SU(6)/Sp(6)$ model in the Higgs vacuum (labelled with their $SU(2)_L \times  U(1)_Y$ quantum numbers) and in the HL vacuum. The $U(1)_X$ charge assignments in the HL vacuum are indicated in the last column. Note that $H_1 = (\omega^+, \frac{h+i z^0}{\sqrt{2}})^T$ and $\omega^+, z^0$ are the Goldstones eaten by $W$ and $Z$.} \label{tab:su6sp6}
\end{table}

How can the vacuum depend on temperature? In general, the potential determining the vacuum alignment has the form (at leading order in the chiral expansion)~\cite{Contino:2010rs}
\beq
V(\theta, T) = - a (T)\ \sin^2 \theta + \frac{1}{2} b(T)\ \sin^4 \theta\,.
\eeq
Assuming $b(T)>0$, the breaking of the EW symmetry can be achieved for $a(T)>0$, with the minimum located at $\sin^2 \theta = a(T)/b(T)$ for $0<a(T)/b(T)<1$ and at $\sin \theta = 1$ for $a(T)/b(T)\geq 1$.  Thus, the vacuum structure needed for our mechanism can be achieved if $a(T)/b(T)$ varies with the temperature and we have:
\beq
\frac{a(T_{\rm HL})}{b(T_{\rm HL})} > 1 \quad  \mbox{and} \quad \frac{a(0)}{b(0)} \ll 1\,,
\eeq
where $T_{\rm HL}$ is identified with the temperature of confinement.
This implies that the above ratio needs to monotonically decrease with temperature, and that the vacuum is stuck at the HL position until the temperature $T_\ast$ for which $a(T_\ast) = b(T_\ast)$. In this period, for $T_{\rm HL} > T > T_{\ast}$, the electroweak breaking scale is $v(T) = f \gg v_{\rm SM}$, and the $W$, $Z$ and SM fermions are much heavier than the SM values by a factor $f/v_{\rm SM}$. If we consider a benchmark scale $f = 1.5$~TeV, this yields $m_W^{\rm HL} (T) = 490$~GeV, $m_Z^{\rm HL} (T) = 560$~GeV, and $m_t^{\rm HL} (T)=1060$~GeV. An example where this thermal dynamics is driven by a composite dilaton has been studied in Refs~\cite{Bruggisser:2018mus,Bruggisser:2018mrt}, while for an example with an elementary Higgs we refer the reader to Refs~\cite{Baldes:2018nel}. A baryon asymmetry can thus be generated at the phase transition via varying Yukawa couplings~\cite{Bruggisser:2018mus} that enhance CP violation with respect to lo energy.

We can now start following the thermal history of the DM candidates. At the phase transition temperature, $T_{\rm HL} \approx \mathcal{O} (f)$, the global symmetry $U(1)_X$ is exact while the EW symmetry is broken. The pNGBs, therefore, can be labelled in terms of their electromagnetic and $U(1)_X$ charges. For the template $SU(6)/Sp(6)$ model, refer to the third column of Table~\ref{tab:su6sp6}. Note that the will-be Higgs $h$ forms a $q_X = 1$ state together with the singlet $\eta$, while all the $\mathbb{Z}_2$--odd pNGBs have charges $q_X = 1/2$. We will call the latter collectively as $\Theta_i$. One interesting point of our model is that $U(1)_X$, together with baryon and lepton numbers, $B$ and $L$ respectively, has an anomaly with the electroweak gauge interactions: if a baryon asymmetry is generated at $T_{\rm HL}$~\cite{Bruggisser:2018mus,Bruggisser:2018mrt} or above (i.e., via high scale Leptogenesis~\cite{Davidson:2008bu}), it will be transferred to an $X$ asymmetry. The relative densities can be computed following Ref.~\cite{Ryttov:2008xe}, with the only difference that we will not include any Higgs boson in the computation as our theory is Higgsless~\footnote{I.e., we assume that the $0^{++}$ state, that may play the role of the Higgs~\cite{Belyaev:2013ida}, is heavy.} at the phase transition. The computation is model-dependent, so here we will show the results for the template model: the relation among the chemical potentials of the various states can be easily computed following their quantum numbers, while we find that the relation imposed by the EW sphalerons is the same as in Ref.~\cite{Ryttov:2008xe},
\beq
2 \mu_\Theta + 9 \mu_{u_L} + 3 \mu_W + \mu_L = 0\,,
\eeq
involving 4 remaining independent chemical potentials: $\mu_\Theta$ of the $\mathbb{Z}_2$--odd pNGBs, $\mu_{u_L}$ of the left-handed up-type quarks, $\mu_W$ of the $W^-$ boson and $\mu_L = \sum_{i=1}^3 \mu_{l^-_i}$ being the total one of the three charged leptons. For a strong 1$^{\rm st}$ order phase transition~\cite{Bruggisser:2018mus}, imposing the vanishing of the total charge and isospin gives a fixed numerical ratio for the asymmetries.
Assuming that all $\mathbb{Z}_2$--odd pNGBs are light compared to the phase transition scale, we find~\footnote{For a 2$^{\rm nd}$ order phase transition, the vanishing of the isospin cannot be imposed and there is one remaining free parameter (identifiable with the $L/B$ ratio), so that we predict $$\frac{X}{B} = - \frac{2 (93+44 \xi)}{63}\,, \;\; \frac{L}{B} = \xi\,.$$}
\beq
 \displaystyle \frac{X}{B} = - 4\,, \quad \frac{L}{B} = \frac{3}{4}\,.
\eeq
We also studied the spectrum of the template model and found that a typical spectrum contains a light pair $\Theta_1$--$\Theta_1^-$ (or $\Theta_2$--$\Theta_2^+$), while the other two are much heavier; we also find that half of the $X$--charge density is initially stored in $\phi_X$, and the other half in the $\mathbb{Z}_2$--odd states $\Theta_i$.

\begin{figure}[tb]
\includegraphics[width=0.4\textwidth]{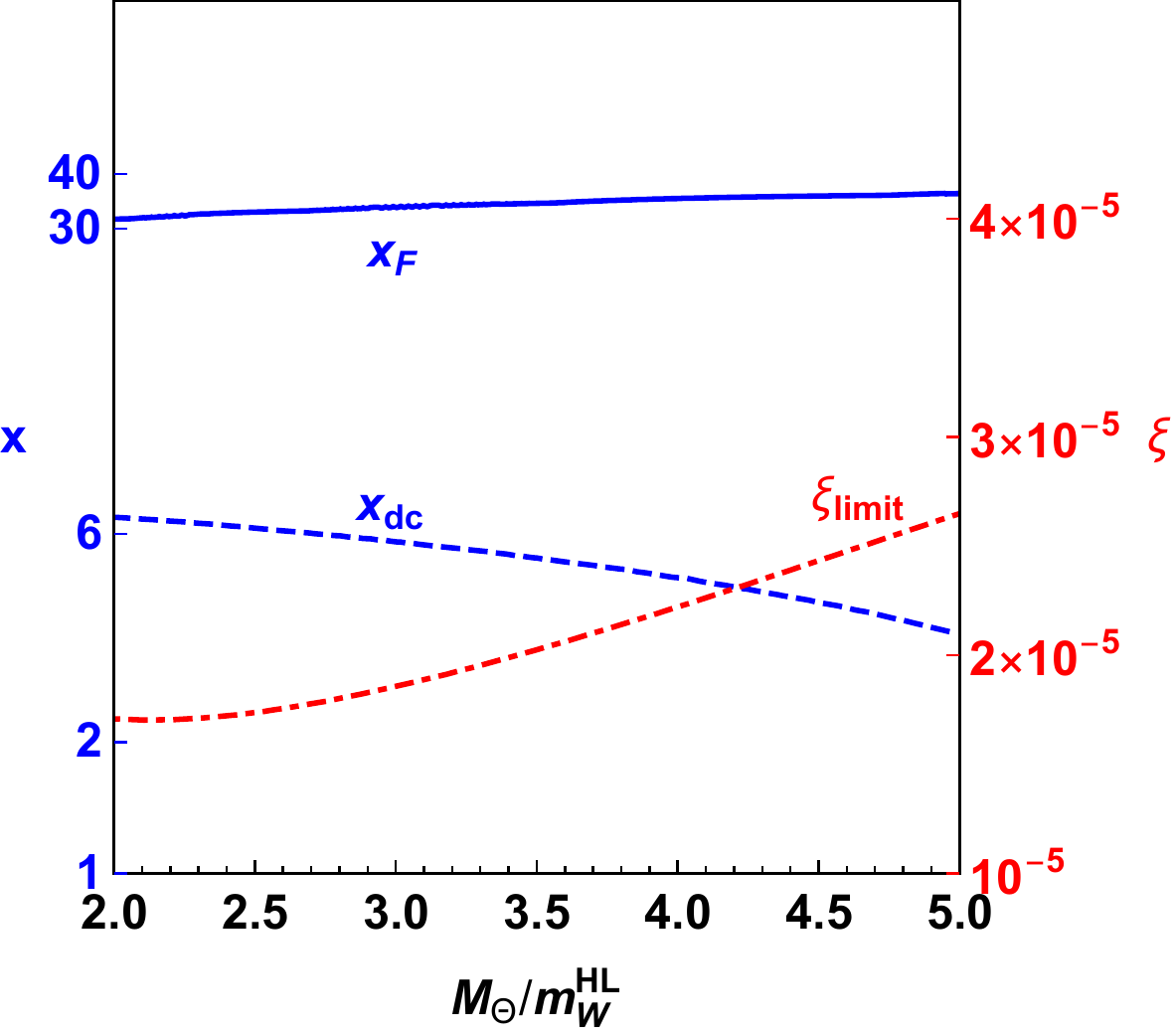}
\caption{In blue, freeze-out $x_F$ and decoupling $x_{dc}$ temperatures for $\Theta_i$ as a function of $M_\Theta/m_W^{\rm HL}$ for $m_W^{\rm HL} = 500$~GeV. In red, the value of $\xi$ from Eq.~(\ref{eq:xi}) that saturates the relic density from the $X$ asymmetry (smaller values are excluded). }
\label{fig:results}
\end{figure}

For the DM density generated by the asymmetry to persist, it is crucial that the $\Theta_i$ states decouple from the thermal bath before the temperature $T_\ast$, while the detailed relic density depends on the processes that determine the equilibrium between the $X$--charged states $\phi_X$ and $\Theta_i$. This dynamics, taking place between $T_{\rm HL}$ and $T_\ast$ is also very model dependent, however all models have similar qualitative features. We study a simplified scenario where only three states are relevant: $\phi_X$, $\Theta$ and $\Theta^-$, with $m_{\Theta} \approx m_{\Theta^-} \approx M_\Theta$, and $f \gg M_{\phi_X} \approx 0$. The latter is justified by the fact that the imaginary part of $\phi_X$, $\eta$, becomes an exact Goldstone at $T = T_{\ast}$. The relevant couplings are:
\begin{multline}
\mathcal{L} \supset - i \frac{g}{\sqrt{2}}\, W^+_\mu\ (\Theta^\ast \overleftrightarrow \partial^\mu \Theta^-) + \frac{\xi}{2} f\ \phi_X^\ast \Theta \Theta + \mbox{h.c.}  \\
- \frac{g^2}{2}\ \phi_X^\ast \phi_X \left( W^+_\mu W^{-,\mu} + \frac{1}{2} Z_\mu Z^\mu \right)\,,
\label{eq:xi}
\end{multline}
where $g$ is the $SU(2)_L$ gauge coupling and $\xi$ is a small $U(1)_X$ conserving coupling generated by the pNGB potential.~\footnote{We neglect the couplings of the $Z$, which gives qualitatively very similar results.} The last term is a relic of the fact that $\phi_X$ contains the will-be Higgs boson, which couples to the massive EW gauge bosons.
The coupling $\xi$ is the only one that transfers $X$ charge between the $\mathbb{Z}_2$--odd states $\Theta_i$ and $\phi_X$, thus the temperature $T_{dc} = M_\Theta/x_{dc}$ where the two decouple and the $X$ charges in $\Theta$ are frozen is determined by this interaction getting off thermal equilibrium: we find that the dominant process is $\Theta^\ast + \phi_X \leftrightarrow \Theta^- + W^+$, whose cross section can be easily computed.
The number density of DM candidates coming from the asymmetry is thus determined by the asymmetry in $\Theta$  at $T_{dc}$, which can be computed by solving the appropriate Boltzmann equation. As decoupling occurs where $\Theta_i$ are non-relativistic, neglecting the contribution of the $W$ chemical potential, we find:
\begin{multline}
\frac{\Omega_\Theta}{\Omega_b} \approx \frac{M_\Theta}{m_p} \left|\frac{X}{B}\right| \frac{\Delta n_{\Theta} (T_{dc})}{n_X (T_{dc})} \approx
\\ \frac{M_\Theta}{m_p}  \times 4 \times e^{-M_\Theta/m_W^{\rm HL}} \times  6 \left( \frac{x_{dc}}{2\pi} \right)^{2} e^{-x_{dc}}\,,
\end{multline}
where $n_X (T_{dc})$ is computed by considering that most of the $X$ asymmetry is stored in $\phi_X$, which is light (relativistic) and in thermal equilibrium, and $m_W^{\rm HL}$ is the $W$ mass in the HL vacuum. Assuming that this is the dominant contribution to the DM relic density today, and that $M_\Theta = M_{\rm DM}$ (the mass of the DM at $T \approx 0$ may be different from $M_\Theta (T_{dc})$), we can thus solve the decoupling and $\Omega_\Theta/\Omega_b \approx 5$ to determine $x_{dc}$ and $\xi$. The result is shown in Fig.~\ref{fig:results} by the dashed-blue and dashed-red lines, for $m_W^{\rm HL} = 500$~GeV, corresponding to $f \approx 1.5$~TeV. We see that $\xi$ is required to be very small, and this is a model-building constraint on the explicit models. In composite models the $\xi$ term breaks the symmetry corresponding to G-parity in QCD and so it is not unreasonable for this to be small. In order for the asymmetry to survive, we need to make sure that $\Theta_i$ decouple from the SM thermal bath before the vacuum moves away from the HL vacuum. The freeze-out temperature $T_F = M_\Theta/x_F$ is determined by the process $\Theta \Theta^\ast \leftrightarrow W^+ W^-$ going out of equilibrium, and the result is shown numerically by the blue solid line in Fig.~\ref{fig:results}.  This implies an upper limit on $T_\ast$:
\beq
T_\ast < T_F \approx \frac{M_\Theta}{33} \approx 30~\mbox{GeV} \times \frac{M_\Theta}{1~\mbox{TeV}}\,.
\eeq
As a final consistency check, we determine the freeze-out temperature for $\phi_X$, coming from the processes $\phi_X \phi_X^\ast \leftrightarrow W^+W^-/ZZ$, giving $T_{\phi} \approx 18~\GeV \times f/(1.5~\mbox{TeV})$: this temperature is below $T_F$ for $M_\Theta \gtrsim m_W^{\rm TC}$.
Note that $\phi_X$ is relativistic at freeze-out.

At the temperature $T_\ast < T_F$, the vacuum of the theory starts drifting away from the HL vacuum. At this time, $U(1)_X$ is spontaneously broken by the vacuum, and $\phi_X$ is no longer protected from decays. Thus, only the fraction of $X$ charge stored in the $\mathbb{Z}_2$--odd pNGBs will survive. The masses of the will-be Higgs $h$ and of $\eta$ split, and $h$ starts acquiring Higgs-like couplings to the SM states, scaling like $\cos \theta$~\cite{Liu:2018vel}. Thus, close to the transition temperature $T_{\ast}$, the couplings are still small and the $W$ and $Z$ bosons heavy. While the Universe cools down, the EW masses gradually decrease to the SM values, while the Higgs couplings approach the SM values as $\cos \theta \to 1$. In our model, therefore, \emph{the $125$--GeV Higgs emerges from the dark} $\phi_X$ state during the relaxation at $T < T_\ast$.
One potential concern is that the $\mathbb{Z}_2$--odd states may re-thermalise with the SM once the theory relaxes to the standard vacuum. From standard WIMP
analysis we know that the decoupling happens at $\tilde{T}/M_{\rm DM} (\tilde{T}) \approx \frac{1}{25}$, which points to temperatures $\tilde{T} > T_F$ if $M_{\rm DM} \approx  M_\Theta$. Thus the DM will not thermalise below $T_F$ even when the SM particles reach their standard masses. Furthermore, the non-derivative coupling to the Higgs $h$, which is related to $\xi$ from Eq.~\eqref{eq:xi}, is small enough to avoid direct detection~\cite{Balkin:2018tma}\footnote{For the favourable value of $\xi$ (C.f. Fig.~\ref{fig:results}) and $M_{\rm DM} = 1$~TeV, we obtain a spin-independent cross section of $\sigma_{\rm SI} \approx 10^{-53} \mbox{cm}^{-2}$, which is 7 orders of magnitude below the current limit and below the reach of future experiments.}, while indirect detection can be avoided if the SM state is dominantly a gauge singlet.
This model can be probed at the LHC and future colliders thanks to the additional $\mathbb{Z}_2$--odd states, like $\Theta^-$, which are produced via their EW couplings.

The DM mechanism we propose has a striking low-energy prediction: the presence of a Goldstone boson associated with the spontaneous breaking of $U(1)_X$ at $T<T_\ast$. This Goldstone is the singlet $\eta$, also emerging from $\phi_X$, and it will eventually acquire a small mass from tiny explicit breaking of $U(1)_X$.
In the minimal composite scenarios, the only linear coupling to SM states is generated by the topological anomaly with the electroweak gauge interactions, given schematically by~\footnote{In elementary realisations, other couplings to SM fermions need to be introduced to allow for $\eta$ decays.}
\beq
\frac{\cos \theta}{f}\ \eta\ \left( g^2 \kappa_W\ W_{\mu\nu} \tilde{W}^{\mu\nu} + {g'}^2 \kappa_B\ B_{\mu \nu} \tilde{B}^{\mu \nu} \right)\,.
\label{eq:WZW}
\eeq
This term contains a coupling to two photons, proportional to $\kappa_{\gamma\gamma} = \kappa_W + \kappa_B$, which is very strongly constrained for $m_\eta \lesssim 1$~GeV (see for instance Ref.~\cite{Bauer:2017nlg}), giving rise to bounds on $f$ many orders of magnitude above the TeV scale. The template model is rather special because it features $\kappa_W = - \kappa_B$
so that $\kappa_{\gamma\gamma} = 0$ at leading order.~\footnote{The same holds for the $SU(4)\times SU(4)/SU(4)$ coset~\cite{Ma:2015gra}.}
Albeit $\eta$ has a photophobic nature~\cite{Craig:2018kne}, couplings to photons and to SM fermions are generated at loop level~\cite{Bauer:2017ris}, thus strong bounds may still arise from astrophysics and cosmology.
In our case, for $m_\eta \lesssim 9$~keV, strong bounds $f > \mathcal{O}(100)$~TeV arise from star evolution~\cite{Raffelt:1994ry,Corsico:2001be,Battich:2016htm}, while for $m_\eta < 100$~MeV, interesting effects may be observed in a future supernova observation if $f$ is in the TeV range~\cite{Raffelt:2006cw}.
The couplings in Eq.~\eqref{eq:WZW} also generate decays $Z \to \gamma \eta$, with $\mbox{BR} = 8 \times 10^{-9}$ in our template model,
which is right below the LEP bound for detector-stable $\eta$~\cite{Akers:1994vh,Acciarri:1997im,Abreu:1996vd} and will be observable at a future $e^+ e^-$ collider.
Bounds from cosmology also apply~\cite{Cadamuro:2011fd}, however a detailed analysis is sensitive to the details of the model and of the cosmological evolution of the theory, and they will be presented elsewhere.

In conclusion, in this letter we have presented a new mechanism for non-thermal DM production via vacuum misalignment. The relic density emerges from an asymmetry at high energies, while the SM-like Higgs boson also emerges from the high-temperature dark sector. The mechanism predicts a light pNGB from the low-temperature breaking of the $U(1)$ symmetry, leading to observable effects in future supernova observations and $Z$ decays at future high-luminosity lepton colliders.

\section*{Acknowledgements}
GC acknowledges partial support from the Labex-LIO (Lyon Institute of Origins) under grant ANR-10-LABX-66 (Agence Nationale pour la Recherche), and FRAMA (FR3127, F\'ed\'eration de Recherche ``Andr\'e Marie Amp\`ere'').
CC and HHZ are supported by the National Natural Science Foundation of China (NSFC) under Grant Nos. 11875327 and 11905300, the China Postdoctoral Science Foundation under Grant No. 2018M643282, the Natural Science Foundation of Guangdong Province under Grant No. 2016A030313313, the Fundamental Research Funds for the Central Universities, and the Sun Yat-Sen University Science Foundation.
MTF and MR acknowledge partial funding from The Council For Independent Research, grant number DFF 6108-00623. The CP3-Origins center is partially funded by the Danish National Research Foundation, grant number DNRF90.
GC, CC and HHZ also acknowledge support from the China-France LIA FCPPL.
GC also thanks the Sun Yat-Sen University for hospitality during the completion of this project.

%
%

\bibliographystyle{JHEP-2-2}

\bibliography{bibHiggsfDark.bib}

\end{document}